\documentclass{article}
\usepackage{amssymb}
\usepackage{amsfonts}
\usepackage{amsmath}
\usepackage{epsfig}
\usepackage{graphicx}

\begin{document}

\title{T violation and the dark sector}
\author{R. Vilela Mendes\thanks{%
e-mail: rvilela.mendes@gmail.com, rvmendes@ciencias.ulisboa.pt;
https://label2.tecnico.ulisboa.pt/vilela/} \\
\textit{CMAFcIO, Universidade de Lisboa, }\\
\textit{\ C6 - Campo Grande, 1749-016 Lisboa}}
\date{ }
\maketitle

\begin{abstract}
It is argued, as a working hypothesis, that "normal" and dark matter
interactions can only be T and CP violating. One way to implement this idea
is to consider that time reversal in dark matter is implemented, not by an
antiunitary operator, but by an unitary operator. It is shown how this
occurs naturally in the context of complex spacetime with an extended
symmetry group.
\end{abstract}

\section{The interactions of dark matter}

There is now extensive evidence for the existence in the Universe of matter
not seen by our instruments. Strong evidence comes from the spiral galaxy
rotation curves measured by visible and radio wavelengths, from the galaxy
clusters matter content being larger than expected from luminosity, from the
virial theorem, from velocity dispersion and gravitational lensing, as well
as from galaxy clusters dynamics, the large scale structure and the cosmic
microwave background radiation anisotropies, etc. Another current
cosmological mystery is the acceleration of the universe expansion, which
may be associated to an essentially constant density of dark energy. This
may or may not be associated to the dark matter phenomena. Or it may simply
mean that Einstein's equation with the cosmological term is the right
classical gravitational equation. In any case matter in the dark sector,
whatever it is, has so far manifested itself only by gravitational
interactions with ordinary matter. What is so special about gravitational
interactions?

A feature that might distinguish gravity from the other interactions at the
quantum level, is the absence of global symmetries. It is of course possible
to construct quantum field theory Lagrangians for matter and gravitational
fields which, at the perturbative level, preserve all symmetries. However it
has for a long time been conjectured \cite{Penrose} \cite{Wald} that, in an
actual quantum theory of gravity, no global symmetries will be preserved.
This conjecture which, in the past, was mostly based on singularity and
black hole physics was somewhat strengthened in the AdS/CFT holographic
context \cite{Harlow}. Admittedly AdS/CFT is not the final quantum gravity
theory, nevertheless it is, so far, the best understood consistent framework.

Among the symmetries that would not be preserved in quantum gravity is
time-reversal (\cite{Harlow}, Sec.7 and Th. 4.2). This suggests as a
possible interpretation (or a working hypothesis) for the elusive nature of
the interactions of dark with ordinary matter, that they must be $T-$%
violating interactions. If $CPT$\ is conserved, at least locally, this might
also imply a $CP-$violating nature for the interactions of dark and ordinary
matter.

That only $T$ and $CP$ violating interactions are associated to dark matter
is a general hypothesis that might be realized in different ways. Here a
particular scenario will be explored, in a complex spacetime context, where
normal and dark matter belong to different superselection sectors, having
different symmetry properties.

\section{T-violation in a complex spacetime scenario}

Several authors have studied extensions of gravitational theory by
complexifying the functions that appear in the classical equations. Already
Einstein in 1945 \cite{Einstein} generalized his theory of gravitation by
considering a metric with complex components and Hermitian symmetry, others 
\cite{Kunstatter} have complexified the tangent bundle over the real
four-dimensional spacetime manifold, constructed solutions of the complex
Einstein equation \cite{Plebanski}, or provided a complex manifold
description of massless particles \cite{Penrose-2} \cite{Esposito-1}, etc.

A more radical approach consisted in assuming that the spacetime manifold
itself should be parametrized but complex coordinates. Several such
explorations have been carried out, see for example \cite{Brown} \cite%
{Newman} and for a more complete set of references and perspectives see Ref.%
\cite{Esposito-2}. Of course, a question that is left open is, if the
spacetime dimensions are indeed indexed by complex numbers, why we do not
feel it in our everyday experience.

Recently this question was revisited \cite{Vilela-2} for spacetimes where
coordinates take values in higher normed division algebras, complex,
quaternion or octonion. In this extended spacetime, any four independent
directions may be chosen as a basis for a real vector space, that is, for a 
\textit{real slice} of the complex spacetime\textit{.} Using as consistency
condition that the symmetry groups of the extended spacetime must reduce to
the real Lorentz and Poincar\'{e} groups in each real slice, one obtains,
for the homogeneous group of the extended spacetime, the condition%
\begin{equation}
\Lambda ^{\dag }G\Lambda =G  \label{2.1}
\end{equation}%
that is, a $U\left( 1,3,\mathbb{K}\right) $ group, $\mathbb{K=C},\mathbb{Q},%
\mathbb{O}$, with $G$ the metric $\left( 1,-1,-1,-1\right) $. For $\mathbb{%
K=C}$ this has been called the complex Lorentz group with real metric \cite%
{Barut}. It is a $16-$parameter group different from the $12-$parameter group%
\begin{equation}
\Lambda ^{T}G\Lambda =G  \label{2.2}
\end{equation}%
used in the analytic continuations of the S-matrix. Nevertheless it is the
condition (\ref{2.1}) that insures an identical group in each one of the
real slices.

The nature of the matter states, that is implied by this group structure,
was obtained by studying the representations of the semidirect group%
\begin{equation}
T_{4}\circledS U\left( 1,3,\mathbb{C}\right)  \label{2.3}
\end{equation}%
$T_{4}$ being the complex spacetime translations. The conclusions were:

- Half-integer spin states are not elementary states for the full complex
group (\ref{2.3}). That is, half-integer elementary states, that are
irreducible representations of the real groups, cannot communicate between
different real slices, in the sense that they cannot be rotated from one
real slice to another.

- Integer-spin states are elementary states of the full group. However these
states have a particular nature when the full group, including reflections,
is taken into account. In the complex Lorentz group (\ref{2.1}) parity and
time reversal are continuously connected to the identity. Therefore, in
faithful continuous norm-preserving representations, both parity and time
reversal must be implemented by unitary operators. On the other hand one
knows that, in each real slice, positivity of energy requires an antiunitary
time reversal operator.

In conclusion: In the complex spacetime scenario there are three types of
states: half-integer spin states that are confined to the real slices,
integer-spin positive energy states that are also only associated to the
real slices and finally integer-spin states, with unitary time reversal
operator, that can communicate between real slices. Because of its
association to time reversal violation, this last type will be denoted here
as \textit{chronoparticles}.

The conclusions are qualitatively similar for the normed division algebras $%
\mathbb{Q}$ and $\mathbb{O}$ \cite{Vilela-2}. Also notice that although the
extended Lorentz group does not contain half-integer elementary states, it
might still operate on these states. However, rather than rotating them
between real slices it generates a multiplicity of states (see \cite%
{Vilela-2}, Appendix D).

It is worthwhile to recall the energy positivity argument that leads to the
antiunitarity of the time reversal operator. When time-reversal is a
conserved symmetry, invariance of the Schr\"{o}dinger equation%
\begin{equation*}
i\partial _{t}\psi =H\psi
\end{equation*}%
implies that either the $T$ operator is antiunitary or it anticommutes with
the Hamiltonian. However, if it is unitary and anticommutes with the
Hamiltonian, then 
\begin{equation*}
\left( T\psi ,HT\psi \right) =-\left( \psi ,H\psi \right)
\end{equation*}%
and there are negative energy states.

Chronoparticles, as defined above, because they connect different real
slices may establish interactions between matter in different real slices.
Because of the argument above, they might behave as negative energy
particles from the point of view of physics in the real slices. However, the
above reasoning does not really apply because $T$ is not conserved in their
interactions.

Because of the unitarity of the $T$ operator, chronoparticles and ordinary
matter in the real slices are in different superselection sectors and in
their interactions $T$ is not a symmetry. The argument \cite{Vilela-2} goes
as follows:

Let $\psi _{o}\in V_{o}$ be a state in the (real slice) ordinary matter
space and $\psi _{c}\in V_{c}$, a chronoparticle (dark matter?) state. There
is a superselection rule operating between these two types of spaces.
Consider a linear superposition of two of these states%
\begin{equation*}
\Phi =\alpha \psi _{o}+\beta \psi _{c}
\end{equation*}%
with $\alpha ,\beta $ real numbers. Now $\Phi $ and $e^{i\theta }\Phi
=\alpha e^{i\theta }\psi _{o}+\beta e^{i\theta }\psi _{c}$ belong to the
same ray and therefore should represent the same state. Applying the time
reversal operator to both $\Phi $ and $e^{i\theta }\Phi $%
\begin{eqnarray}
T\Phi &=&\alpha T\psi _{o}+\beta T\psi _{c}  \notag \\
Te^{i\theta }\Phi &=&\alpha e^{-i\theta }T\psi _{o}+\beta e^{i\theta }T\psi
_{c}  \label{2.5}
\end{eqnarray}%
$T\Phi $ and $Te^{i\theta }\Phi $ belong to different rays, hence $T$ does
not establish a ray correspondence in $V_{o}\oplus V_{c}$ unless $\alpha =0$
or $\beta =0$, that is, $V_{o}$ and $V_{c}$ belong to different
superselection sectors.

The fact that the states are in different superselection sectors does not
mean that they cannot interact. Whereas the superselection rule result
concerns the structure of the direct sum $V_{o}\oplus V_{c}$, the nature of
the interactions depends on the way the group transformations operate in the
tensor product $V_{o}\otimes V_{c}$. Consider now the $T$ operation acting
on $V_{o}\otimes V_{c}$ and compute its action on a matrix element%
\begin{equation}
\left( T\left( \psi _{o}^{(1)}\otimes \psi _{c}^{(1)}\right) ,T\left( \psi
_{o}^{(2)}\otimes \psi _{c}^{(2)}\right) \right) =\left( U_{o}\psi
_{o}^{(2)}\otimes U_{c}\psi _{c}^{(1)},U_{o}\psi _{o}^{(1)}\otimes U_{c}\psi
_{c}^{(2)}\right)  \label{2.6}
\end{equation}%
where $U_{o}$ and $U_{c}$ are unitary operators, acting on the degrees of
freedom of the states. From (\ref{2.6}) one concludes that there is no
choice of phases that can make $T$ a global unitary or anti-unitary operator
in the tensor product space. Therefore, by Wigner's theorem, $T$ cannot be a
symmetry in $V_{o}\otimes V_{c}$.

How a particle with a unitary implementation of the $T$ operation may induce
symmetry violations in its interactions with ordinary matter is simple to
illustrate. Let the Fourier decompositions of the quantum fields $\Phi _{o}$
and $\Phi _{c}$, associated to ordinary matter and to an integer-spin
chronoparticle be%
\begin{eqnarray}
\Phi _{o}\left( x^{0},\overrightarrow{x}\right) &=&\sum_{\mu }\int d^{3}k%
\frac{\epsilon \left( k,\mu \right) }{2\omega \left( k\right) }\left\{
a_{o}\left( k,\mu \right) e^{-ik.x}+b_{o}^{\dag }\left( k,\mu \right)
e^{ik.x}\right\}  \notag \\
\Psi _{c}\left( x^{0},\overrightarrow{x}\right) &=&\sum_{\mu }\int d^{3}k%
\frac{\epsilon \left( k,\mu \right) }{2\omega \left( k\right) }\left\{
a_{c}\left( k,\mu \right) e^{-ik.x}+b_{c}^{\dag }\left( k,\mu \right)
e^{ik.x}\right\}  \label{2.7}
\end{eqnarray}%
$\mu $ being an helicity label and $\epsilon \left( k,\mu \right) $ the
polarization vectors.

With%
\begin{eqnarray}
P\Phi \left( x^{0},\overrightarrow{x}\right) P^{-1} &=&\eta _{P}\Phi \left(
x^{0},-\overrightarrow{x}\right)   \notag \\
T\Phi \left( x^{0},\overrightarrow{x}\right) T^{-1} &=&\eta _{T}\Phi \left(
-x^{0},\overrightarrow{x}\right)   \notag \\
C\Phi \left( x^{0},\overrightarrow{x}\right) C^{-1} &=&\eta _{C}\Phi ^{\dag
}\left( x^{0},\overrightarrow{x}\right)   \label{2.8}
\end{eqnarray}%
for both ordinary matter and chronoparticle fields, one obtains%
\begin{eqnarray}
Pa_{o}\left( k^{0},\overrightarrow{k}\right) P^{-1} &=&\eta _{P}a_{o}\left(
k^{0},-\overrightarrow{k}\right)   \notag \\
Pb_{o}^{\dag }\left( k^{0},\overrightarrow{k}\right) P^{-1} &=&\eta
_{P}b_{o}^{\dag }\left( k^{0},-\overrightarrow{k}\right)   \notag \\
Ta_{o}\left( k^{0},\overrightarrow{k}\right) T^{-1} &=&\eta _{T}a_{o}\left(
k^{0},-\overrightarrow{k}\right)   \notag \\
Tb_{o}^{\dag }\left( k^{0},\overrightarrow{k}\right) T^{-1} &=&\eta
_{T}b_{o}^{\dag }\left( k^{0},-\overrightarrow{k}\right)   \notag \\
Ca_{o}\left( k^{0},\overrightarrow{k}\right) C^{-1} &=&\eta _{C}b_{o}\left(
k^{0},\overrightarrow{k}\right)   \notag \\
Cb_{o}^{\dag }\left( k^{0},\overrightarrow{k}\right) T^{-1} &=&\eta
_{C}a_{o}^{\dag }\left( k^{0},\overrightarrow{k}\right)   \label{2.9}
\end{eqnarray}%
for ordinary matter, with $P$ and $C$ unitary and $T$ antiunitary and%
\begin{eqnarray}
Pa_{c}\left( k^{0},\overrightarrow{k}\right) P^{-1} &=&\eta _{P}a_{c}\left(
k^{0},-\overrightarrow{k}\right)   \notag \\
Pb_{c}^{\dag }\left( k^{0},\overrightarrow{k}\right) P^{-1} &=&\eta
_{P}b_{c}^{\dag }\left( k^{0},-\overrightarrow{k}\right)   \notag \\
Ta_{c}\left( k^{0},\overrightarrow{k}\right) T^{-1} &=&\eta _{T}b_{c}^{\dag
}\left( k^{0},-\overrightarrow{k}\right)   \notag \\
Tb_{c}^{\dag }\left( k^{0},\overrightarrow{k}\right) T^{-1} &=&\eta
_{T}a_{c}\left( k^{0},-\overrightarrow{k}\right)   \notag \\
Ca_{c}\left( k^{0},\overrightarrow{k}\right) C^{-1} &=&\eta _{C}a_{c}^{\dag
}\left( k^{0},\overrightarrow{k}\right)   \notag \\
Cb_{c}^{\dag }\left( k^{0},\overrightarrow{k}\right) T^{-1} &=&\eta
_{C}b_{c}\left( k^{0},\overrightarrow{k}\right)   \label{2.10}
\end{eqnarray}%
for chronoparticle operators, with $P$ and $T$ unitary and $C$ antiunitary
(for $CPT$ invariance). For simplicity the helicity labels were dropped.

Now it is clear in which way chronoparticles violate $T-$invariance. Let the
interaction term of ordinary matter with chronoparticles be%
\begin{equation*}
\mathcal{L}_{I}\sim \Phi _{o}^{\dag }\Phi _{o}\Phi _{c}
\end{equation*}%
Then, the simple vertex%
\begin{equation*}
\left\langle \phi _{o\left( \overrightarrow{k}+\overrightarrow{v}\right)
}\left\vert a_{o}^{\dag }\left( \overrightarrow{k}+\overrightarrow{v}\right)
a_{c}\left( \overrightarrow{v}\right) a_{o}\left( \overrightarrow{v}\right)
\right\vert \phi _{o\left( \overrightarrow{k}\right) }\psi _{c\left( 
\overrightarrow{v}\right) }\right\rangle
\end{equation*}%
where an ordinary matter state $\phi _{o\left( \overrightarrow{k}\right) }$
absorbs a chronoparticle $\psi _{c\left( \overrightarrow{v}\right) }$
leading to $\phi _{o\left( \overrightarrow{k}+\overrightarrow{v}\right) }$,
becomes, under the $T-$operation%
\begin{equation*}
\left\langle \phi _{o\left( -\overrightarrow{k}\right) }\psi _{c\left( -%
\overrightarrow{v}\right) }^{\dag }\left\vert a_{o}^{\dag }\left( -%
\overrightarrow{k}\right) b_{c}^{\dag }\left( -\overrightarrow{v}\right)
a_{o}\left( -\overrightarrow{k}-\overrightarrow{v}\right) \right\vert \phi
_{o\left( -\overrightarrow{k}-\overrightarrow{v}\right) }\right\rangle
\end{equation*}%
rather than%
\begin{equation*}
\left\langle \phi _{o\left( -\overrightarrow{k}\right) }\psi _{c\left( -%
\overrightarrow{v}\right) }\left\vert a_{o}^{\dag }\left( -\overrightarrow{k}%
\right) a_{c}^{\dag }\left( -\overrightarrow{v}\right) a_{o}\left( -%
\overrightarrow{k}-\overrightarrow{v}\right) \right\vert \phi _{o\left( -%
\overrightarrow{k}-\overrightarrow{v}\right) }\right\rangle
\end{equation*}

That is, the $T-$operation leads to a process different from the usual
time-reversed one and therefore does not establish an identification of the
couplings for the direct and the reversed processes. Notice that in this
example, because $C$ is chosen to be antiunitary for chronoparticles, one
has $CT$ and $P$ invariance but $CP$ and $T$ violation. Chronoparticles may
induce $T-$violating interactions, either by establishing interactions
between matter particles in different real slices or in each real slice by
virtual chronoparticle interactions.


\begin{thebibliography}{99}
\bibitem{Penrose} R. Penrose; \textit{Singularities and time-asymmetry}, in 
\textit{General Relativity: An Einstein Centenary Survey}, S. W. Hawking and
S. W. Israel (Eds.), pp. 581-638, Cambridge U. P., Cambridge 1979.

\bibitem{Wald} R. M. Wald; \textit{Quantum gravity and time reversibility},
Phys. Rev. D 21 (1980) 2742-2755.

\bibitem{Harlow} D. Harlow and H. Ooguri; \textit{Symmetries in Quantum
Field Theory and Quantum Gravity}, Comm. Math. Phys. 383 (2021) 1669--1804.

\bibitem{Einstein} A. Einstein; \textit{A generalization of the relativistic
theory of gravitation}, Annals of Math. 46 (1945) 578-584.

\bibitem{Kunstatter} G. Kunstatter and R. Yates; \textit{The geometrical
structure of a complexified theory of gravitation}, J. Phys. A: Math Gen. 14
(1981) 847-854.

\bibitem{Plebanski} J. Plebanski; \textit{Some solutions of complex Einstein
equations}, J. Math. Phys. 16 (1975) 2395-2402.

\bibitem{Penrose-2} R. Penrose; \textit{The twistor approach to space-time
structures}, in \textit{100 Years of Relativity}, A. Ashtekar (Ed.) pp.
465-505, World Scientific, Singapore 2005.

\bibitem{Esposito-1} G. Esposito; \textit{From spinor geometry to complex
general relativity, }Int. J. of Geometric Methods in Modern Physics 2 (2005)
675-731.

\bibitem{Brown} E. H. Brown; \textit{On the complex structure of the universe%
}, J. Math. Phys. 7 (1966) 417-425.

\bibitem{Newman} E. T. Newman; \textit{Maxwell's equations and complex
Minkowski spac}e; J. Math. Phys. 14 (1973) 102-103.

\bibitem{Esposito-2} G. Esposito; C\textit{omplex geometry of Nature and
general relativity}, Kluwer Acad. Press, Dordrecht 2002.

\bibitem{Vilela-2} R. Vilela Mendes; \textit{Space times over normed
division algebras, revisited}, Int. J. Modern Physics A 35 (2020) 2050055.

\bibitem{Barut} A. O. Barut; \textit{Complex Lorentz group with a real
metric: Group structure}, J. Math. Phys. 5 (1964) 1652-1656.
\end{thebibliography}
\end{document}